\documentclass{article}

\usepackage{arxiv}

\usepackage[utf8]{inputenc} 
\usepackage[T1]{fontenc}    
\usepackage{hyperref}       
\usepackage{url}            
\usepackage{booktabs}       
\usepackage{amsfonts,amsmath}       

\usepackage{lipsum}		
\usepackage{graphicx}
\usepackage{caption}
\usepackage{subcaption}
\usepackage{float}

\title{Predicting Multi-Joint Kinematics of the Upper Limb from EMG Signals Across Varied Loads with a Physics-Informed Neural Network}

\author{
  Rajnish Kumar\\
  Department of Applied Mechanics \\
  Indian Institute of Technology \\
  New Delhi\\
  \texttt{Rajnish.Kumar@am.iitd.ac.in}\\
   \And
  Suriya Prakash Muthukrishnan \\
  Department of Physiology \\
  All India Institute of Medical Sciences \\
  New Delhi\\
  \texttt{dr.suriyaprakash@aiims.edu} \\
   \And
  Lalan Kumar \\
  Department of Electrical Engineering \\
  Indian Institute of Technology\\
  New Delhi\\
  \texttt{lalank@ee.iitd.ac.in} \\
   \And
  Sitikantha Roy \\
  Department of Applied Mechanics \\
  Indian Institute of Technology\\
  New Delhi\\
  \texttt{sroy@am.iitd.ac.in} \\
}

\date{}

\renewcommand{\shorttitle}{Predicting Multi-Joint Kinematics from EMG Signals with a PINN}

\begin{document}
\maketitle

\begin{abstract}
In this research, we present an innovative method known as a physics-informed neural network (PINN) model to predict multi-joint kinematics using electromyography (EMG) signals recorded from the muscles surrounding these joints across various loads. The primary aim is to simultaneously predict both the shoulder and elbow joint angles while executing elbow flexion-extension (FE) movements, especially under varying load conditions. The PINN model is constructed by combining a feed-forward Artificial Neural Network (ANN) with a joint torque computation model. During the training process, the model utilizes a custom loss function derived from an inverse dynamics joint torque musculoskeletal model, along with a mean square angle loss. The training data- set for the PINN model comprises EMG and time data collected from four different subjects. To assess the model’s performance, we conducted a comparison between the predicted joint angles and experimental data using a testing data set. The results demonstrated strong correlations of 58\% to 83\% in joint angle prediction. The findings highlight the potential of incorporating physical principles into the model, not only increasing its versatility but also enhancing its accuracy. The findings could have significant implications for the precise estimation of multi-joint kinematics in dynamic scenarios, particularly concerning the advancement of human-machine interfaces (HMIs) for exoskeletons and prosthetic control systems.
\end{abstract}


\section{Introduction}
The electromyography (EMG), which reflects the neuromuscular activities of humans, serves as a direct source for deciphering human motor intentions. It has been extensively employed in the development of human-machine interfaces (HMIs) that utilize exoskeletons or prostheses to enhance strength and endurance \cite{farina2017man}\cite{koller2015learning}. Decoding the intention to activate multiple degrees of freedom (DOFs) in a coordinated manner and accounting for variations in loading to replicate natural human movements poses a significant challenge within the field of EMG-based HMIs \cite{muceli2011simultaneous} \cite{jung2021intramuscular}. To capture natural human movement accurately, it's essential to measure limb trajectory. While driving an exoskeleton trajectory data enables us to monitor parameters such as range of motion and speed, during task performance. 

Physiological model-based approaches have been utilized to create HMIs by connecting EMG signals to joint kinematic characteristics through EMG-driven musculoskeletal models. An approach combining a geometric model with a musculoskeletal model has been proposed for the continuous motion estimation of the elbow joint \cite{pau2012neuromuscular}. Musculoskeletal (MSK) models are used for real-time simulation of shoulder and elbow joint movements, incorporating passive dampers to avoid numerical stiffness issues \cite{chadwick2008real}. Additionally, an implicit musculoskeletal model formulation has been introduced to control wrist and hand movements in real-time \cite{blana2017real} \cite{crouch2016lumped}. These models are constructed based on biomechanical principles and anatomical data to replicate how muscles, bones, and joints interact.

Despite providing reasonably accurate estimations, musculoskeletal models have notable drawbacks. They are computationally complex and require customization to individual subjects, involving the estimation of numerous parameters. This parameter estimation process is challenging and time-consuming. Furthermore, in real-time applications demanding quick and immediate responses, the computational demands of these models can hinder their effectiveness.

Over the past few years, engineering applications have witnessed a growing adoption of machine learning (ML) and deep learning techniques. This trend is driven by their capacity to uncover intricate features and patterns within data. In the domain of human motion prediction, data-driven approaches have emerged, establishing direct connections between sEMG signals and joint kinetics/kinematics. This eliminates the necessity for forward dynamics equations and parameter estimation, as referenced in previous studies \cite{au2000emg,cote2019deep,ma2020continuous, hajian2022deep}.

Although ML-based approaches offer the advantage of eliminating the need for calibrating physiological parameters and are quicker to implement compared to musculoskeletal models, they do come with specific limitations. These methods operate as 'black boxes,' relying on a generic mapping function without explicitly defining the functional connections between neural commands and the corresponding movements. Achieving successful predictions from given EMG signals necessitates a substantial dataset that includes both EMG signals and their associated motions for training the transfer function. Furthermore, data-driven approaches often struggle to generalize effectively to various tasks, resulting in accuracy issues when there are variations in movements or actions, as highlighted in a previous study \cite{sartori2016neural}. Consequently, these models lack transparency and may not completely align with the fundamental principles of physics, which can be a limitation in specific applications.

In more recent developments, a novel technique known as physics-informed neural networks (PINNs) has gained prominence, as referenced in previous works \cite{raissi2019physics, karniadakis2021physics}. PINNs are engineered to approximate solutions to particular physical equations through the use of feed-forward neural networks while reducing disparities in the governing partial differential equations (PDEs) and their related initial and boundary conditions. In the domain of biomechanics and biomedical applications, the PINN approach has demonstrated success, as evidenced in previous works \cite{zhang2022physics, kissas2020machine}. Furthermore, it exhibits strong performance when applied to solving inverse problems, as indicated by the study \cite{taneja2022feature}.

In this work, a physics informed neural network (PINN) is explored for the first time in the human upper extremity  to predict multi-joint kinematics at different movement conditions using EMG signals. In particular, PINN is utilized for the upper extremity that can estimate joint angles at the shoulder and elbow concurrently for a task that involves flexion-extension (FE) of the elbow at various loading conditions. The PINN is trained across multi-joint to a person using appropriate physiological criteria governed by the dynamics of motion. As a result, we tested the trained model against the accuracy of the predicted joint kinematics. This will provide us with the assurance that we can estimate multiple joint angles from EMG signals of the muscles spanning those joints. This finds application in the creation of a reliable data-driven HMI for the simultaneous and proportional control of numerous DOFs in upper extremity wearable robotic systems.

The rest of this article is organized as follows. The architecture of feed forward artificial neural network (ANN) and PINN are introduced in Section II. Section III provides the training and the system validation procedures. Section IV presents the results. Discussion is presented in Section V. Section VI concludes this article.

\section{Methodology}
First, we begin by describing the mathematical formulations for the fundamental structure of the proposed physics-informed neural network model. This model employs a feedforward artificial neural network and a customized loss function with the goal of predicting multi-joint angles based on surface electromyography signals. Following this, we delve into the intricacies of the artificial neural network's architecture design, the process for estimating joint torque and the PINN training procedure.

\subsection {Feed forward Artificial Neural Network (ANN)}
The multi-layer artificial neural network model utilized for the estimation of joint kinematics is composed of a dense input layer, which accepts experimentally recorded data as input variables, a dense output layer responsible for the prediction of joint angles, and three dense hidden layers that are instrumental in facilitating the convergence of the neural network.

Artificial Neural Networks (ANNs) operate by processing information in a unidirectional fashion, sequentially passing through input neurons, hidden neurons, and output neurons, as depicted in ANN block Figure 1. The feed-forward process, also known as the forward pass, can be elucidated as follows:

Within the input layer, each neuron represents input variables and accepts discrete EMG signals and discrete time data ${X} \in \mathbb{R}^{(m+1)}$ as input. Here, $m$ corresponds to the number of muscles from which EMG data has been recorded. This input is subsequently utilized to establish connections with discrete joint kinematics, denoted as ${\boldsymbol{q}} \in \mathbb{R}^J$, where $J$ represents the number of joints under consideration. This connection is facilitated through the application of a nonlinear activation function denoted as $\sigma (\cdot)$.
The input variables are then passed through the hidden layers using the following propagation rule:

\begin{equation}
    \boldsymbol{z}^{(i)} = \boldsymbol{W}^{(i)} \boldsymbol{a}^{(i-1)} + \boldsymbol{b}^{(i)}, \quad i = 1, \ldots, L
\end{equation}

Additionally, the activation is computed as:

\begin{equation}
    \boldsymbol{a}^{(i)} = \sigma\left(\boldsymbol{z}^{(i)}\right), \quad i = 1, \ldots, L
\end{equation}

Here, $L$ signifies the number of hidden layers, with the superscript $i$ indicating the specific layer number. The activation $\boldsymbol{a}^{(i)} \in \mathbb{R}^{(n_k)}$ represents the output of layer $i$, where $n_k$ is the number of neurons in that layer.

In the context of the input layer, $n_0 = (m+1)$ denotes the input dimension, and $\boldsymbol{a}^{(0)} = X$, where $X$ is the input data. The trainable parameters encompass $\boldsymbol{W}^{(i)} \in \mathbb{R}^{n_k \times n_{k-1}}$ and $\boldsymbol{b}^{(i)} \in \mathbb{R}^{n_k}$, which correspond to the weights and biases, respectively. The complete set of trainable parameters for the artificial neural network is represented as $\boldsymbol{\theta} = \left\{\boldsymbol{W}, \boldsymbol{b}\right\}$, where $\boldsymbol{W} = \boldsymbol{W}^{(i)}$ and $\boldsymbol{b} = \boldsymbol{b}^{(i)}$, here $\quad i = 1,\ldots, L$, encompassing all weights and biases.

In the hidden layers, a hyperbolic tangent (tanh) activation function, denoted as $\sigma(\cdot)$, is utilized. This activation function is applied to all components of its input vector. However, for the output layer, a sigmoid activation, represented as $\sigma(\boldsymbol{z}) \in (0, 1)$, is employed. This choice ensures that the output of the final hidden layer is mapped to the prediction range between 0 and 1 in the output layer. The resulting output, denoted as $\boldsymbol{\hat{q}}$, represents the approximation of the training kinematic data $\boldsymbol{q}$. It is computed as:

\begin{equation}
\boldsymbol{\hat{q}} = \boldsymbol{W}^{(L+1)} \boldsymbol{a}^{(L)} + \boldsymbol{b}^{(L+1)}
\end{equation}

An ANN serves as a surrogate model for predicting the forward dynamics of musculoskeletal (MSK) systems. Its main objective is to estimate joint kinematics by analyzing past electromyography (EMG) signal data derived from various muscle groups. In this context, the extensive input for motion prediction is derived from the EMG signals, encompassing all the intricate details of the EMG signals recorded by sensors.

During the training process of the ANN model, the mean square error (MSE) between the predicted and measured joint angles is employed as the loss function, denoted as $L_q$, as depicted in the following equation:

\begin{equation}\label{Equation data loss}
L_{q}=\sum_l^{N_{\text {load }}}\sum_t^{N_{\text {trials }}} \sum_d^{N_{\text {dof }}} \text { Error }_{l, t, d} \\
\end{equation}

\begin{equation}\label{Equation data error}
    \operatorname{Error}_{l, t, d}=\frac{1}{N_r} \sum_{r=1}^{N_{data}}\left\|\hat{\boldsymbol{q}}_r\left({X}_r ; \boldsymbol{\theta}\right)-\boldsymbol{q}_r\right\|_{L_2}^2
\end{equation}
Here, the notation $\left\|(\cdot)\right\|_{L_2}^2$ refers to the square of the L2 norm. The variable $r$ signifies the number of data samples in the $t^{th}$ trial at a $d^{th}$ joint under the $l^{th}$ loading condition. $\boldsymbol{\hat{q}}$ represents the predicted joint angle, while $\boldsymbol{q}$ corresponds to the actual angle for the $r^{th}$ sample.
\begin{figure}[h]
  \begin{center}
  \includegraphics[scale=0.4]{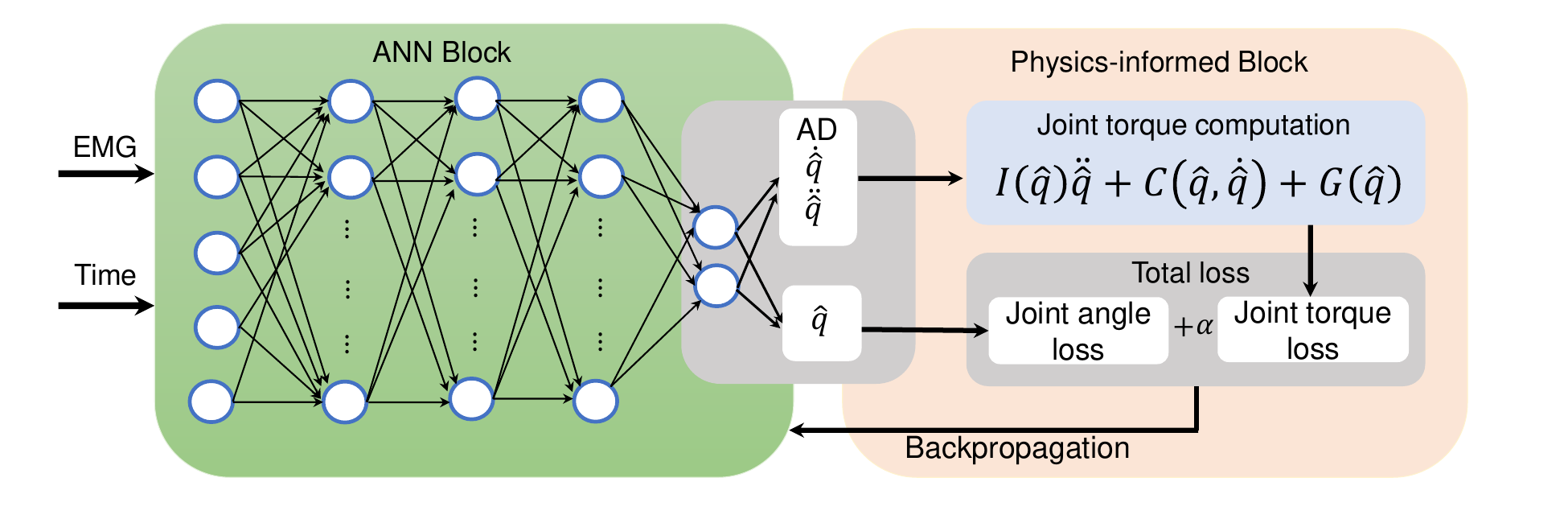}\\
  \caption{The training framework of physics-informed neural network. Here EMG and Time are the input to the ANN block and Joint angle $\hat{q}$ is the output. A predicted joint angle is used to compute joint torque.   }\label{PINN_diagram}
  \end{center}
\end{figure}

\subsection {Physics informed neural network}
The fundamental framework of the suggested deep learning method that combines physics to decipher multi-joint kinematics is presented in Figure \ref{PINN_diagram}. This approach is specifically designed for predicting joint kinematics based on surface electromyography (EMG) data. For a more detailed explanation, in the data-driven segment, an artificial neural network (ANN) is employed to independently acquire an understanding of the forward dynamics of musculoskeletal (MSK) systems. It establishes a connection between EMG signals and joint movement. Simultaneously, the physics-informed part involves the essential physical linkage between joint motion and joint torque.

In this approach, the data-driven stage begins by feeding the recorded EMG signals and their corresponding time intervals into an artificial neural network (ANN). Leveraging the musculoskeletal characteristics acquired by the ANN, predictions can be made for joint angles across various joints under different loading conditions. These predictions must align with the constraints imposed by the equations of motion, which are introduced as adaptable constraints to modify the ANN's loss function. As a result, a customized overall loss function is created for the training procedure by merging the standard mean square error (MSE) loss with the physics-based loss component.

\subsection {Inverse Dynamics}
\begin{figure}[ht!]
	\centering
	\centerline{\includegraphics[scale=0.4]{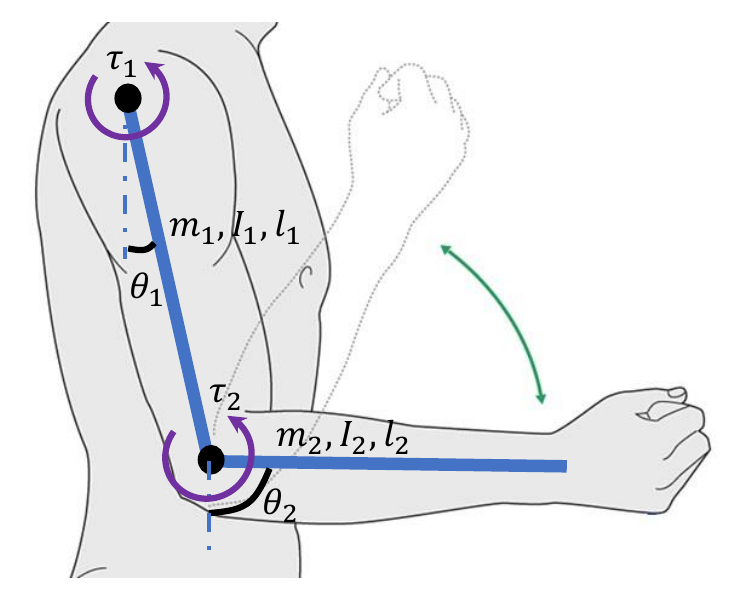}}
	\caption{Double-link model of upper limb.}
	\label{Figure double link model}
\end{figure}

The inverse dynamics method is a traditional approach used to estimate the joint torques and forces in a 2-dimensional (or 3-dimensional) musculoskeletal system. In the case of elbow flexion-extension (FE) motion in the sagittal plane, it involves two degrees of freedom (DOF), one at the shoulder joint and the other at the elbow Figure. \ref{Figure double link model}. Both of these joints can be assumed to be revolute joints, as discussed in previous studies \cite{london1981kinematics,holzbaur2005model}. 
The motion involves two segments: the upper arm (humerus) and the lower arm (ulna and radius, which are treated as a rigid segment). In the inverse dynamics approach, joint torques and forces are calculated based on the motion's kinematics and a representative model of the subject. To illustrate, the positions of the upper limb segments can be captured using markers with a camera-based video system. These data can then be differentiated to determine the velocities and accelerations of the moving limb. The accelerations, velocities, and the physical characteristics of the body (such as mass and inertia) are used as inputs in the equations of motion to calculate the corresponding joint torques and forces. The mass and the center of mass of the body segment can be estimated as described in \cite{herman2016physics} whereas the segment moment of inertia (MOI) can be estimated by the regression equation given Hinrichs et al. \cite{hinrichs1985regression} . The general form of the equation of motion can be given as equation \ref{Equation of motion},
\begin{equation}\label{Equation of motion}
    I({\boldsymbol{q(t)}}) {\boldsymbol{\ddot{q}(t)}}+{{C(\boldsymbol{q(t)}}}, {\boldsymbol{\dot{q}(t)}})+{{G(\boldsymbol{q(t)})}}=\boldsymbol{{\tau(t)}}
\end{equation}
where $q(t)$, ${\dot{q}}(t)$, ${\ddot{q}}(t)$ are the vectors of generalized coordinates, angular velocities, and angular accelerations, respectively; $I({q(t)})$ is the system inertia matrix; $C({q(t)},{\dot{q(t)}})$ is the centrifugal and Coriolis force matrix; $G({q(t)})$ gravitational force on the system. Given the system generalized angular motions $q(t)$, velocity ${\dot{q}}(t)$, accelerations ${\ddot{q}}(t)$, system matrix $I$,$C$ and $G$ joint torques ${\tau}(t)$ can be computed.

\subsection {Customized Loss Function}
In addition to training the ANN to learn the forward dynamics, the suggested framework seeks to direct the training process by minimizing the loss related to the governing equation $\operatorname{L}_{\tau}$ of the musculoskeletal system dynamics, as represented in Equation \ref{Equation of motion}.
\begin{equation}\label{equation tau loss}
    \operatorname{L}_{\tau}=\frac{1}{N_r} \sum_{r=1}^{N_{data}}\left\|\boldsymbol{f}\left(\hat{\boldsymbol{q}}_r\left({X}_r ; \boldsymbol{\theta}\right)\right)\right\|_{L_2}^2
\end{equation}
\begin{equation}\label{equation tau rsidual}
\boldsymbol{f}\left(\hat{\boldsymbol{q}}_r\left({X}_r ; \boldsymbol{\theta}\right)\right) = I(\hat{\boldsymbol{{q}_r}}) {\boldsymbol{\ddot{\hat{q_r}}}}+{{C({\hat{\boldsymbol{q_r}}}}}, {\boldsymbol{\dot{\hat{q_r}}}})+{{G({\hat{\boldsymbol{q_r}}})}}-\boldsymbol{{\tau}}
\end{equation}
Here, $\boldsymbol{f}\left(\hat{\boldsymbol{q}}_r\left({X}_r ; \boldsymbol{\theta}\right)\right)$ is the part represents the vector of discrepancies or residuals related to Equation \ref{equation tau rsidual} for the $r^{th}$ sample.. $\left(\hat{\boldsymbol{q}}_r\left({X}_r ; \boldsymbol{\theta}\right)\right)$ component refers to the vector of joint angles predicted by the ANN with trainable parameters $\boldsymbol{\theta}$ for the $r^{th}$ sample. The best-fitting ANN parameters, represented as $\boldsymbol{\hat{\theta}}$, are obtained by minimizing the custom-designed loss function $J$ as shown below:
\begin{equation}\label{equation min loss}
\hat{\boldsymbol{\theta}}=\underset{\boldsymbol{\theta}}{\operatorname{argmin}}(J)
\end{equation}
\begin{equation}\label{equation custom loss}
\operatorname{J} = L_{q} + \alpha L_{\tau}
\end{equation}
In this context, $\alpha$ is a hyperparameter that regulates the relative significance of the loss contribution stemming from the ODE residual term in the combined loss function. It is determined manually during the training process. In this study, $\alpha$ is adjusted to ensure that the terms $L_{\tau}$ and $L_{q}$ in the loss function, as presented in Equation \ref{equation custom loss}, have the same units. The gradients of the ANN outputs concerning time (input) can be efficiently acquired through automatic differentiation (AD), as depicted in Figure \ref{PINN_diagram}.

\subsection {Architecture and Training of PINN}
This work showcases the effectiveness of a physics-informed machine learning framework. It begins by defining the overall loss within the context of the Physics-Informed Neural Network (PINN) architecture. To minimize this loss, an adaptive optimization method known as Adam is used. Adam dynamically tunes the parameters ($\boldsymbol{\theta}$) of the neural network during the training phase. This adjustment enables the network to learn the forward dynamics of a MSK system and make predictions about joint kinematics while simultaneously adhering to the differential equation constraint imposed by the MSK system's dynamics. This process essentially creates a Physics-Informed Neural Network.

The architecture used in this framework consists of an input layer, which represents the features or attributes of the input data. Hidden layers follow the input layer, where data is processed and non-linearity is introduced to the model. The model's output layer generates predictions or results based on computations performed by neurons in the hidden layers.

The neural network employed in this study is fully connected and comprises four hidden layers, with each layer containing 75 neurons. The activation function used for each neuron in the hidden layers is the hyperbolic tangent function (represented as $\sigma = \tanh(\cdot)$), while the Sigmoid function is applied at the output layer.

To train the network effectively, the Adam optimizer with momentum is applied using mini-batches. Both the data and the physics are processed in batches of 75. The network undergoes training for a total of 1000 epochs, and in this setup, a step size of 300 is specified. This implies that after every 300 epochs, the learning rate will be adjusted. The reduction factor (gamma) is established at 0.8, which means that the learning rate will be multiplied by 0.8 at each step. This approach allows the model to start with a moderately high initial learning rate, aiding rapid convergence during the initial stages of training. As training progresses, the learning rate decreases, helping the model fine-tune its parameters and approach a more optimal solution.

\section{SYSTEM VALIDATION PROCEDURES}
\subsection {Experimental setup, Data collection and Processing}
The experiment involved four healthy (26$\pm$3) year-old male participants with a body mass of (74 $\pm$ 4) kg and height of (172 $\pm$ 3) cm. The participants were directed to execute biceps curls using weights of 0 kg, 2 kg, and 4 kg. Throughout the movement, both the elbow joint and the shoulder joint were permitted to move in the sagittal plane. Simultaneous recording of electromyography (EMG) and joint angle data was conducted. The experimental protocol was approved by the All India Institute of Medical Science (AIIMS) in New Delhi. EMG data was captured using Noraxon’s wireless Ultium EMG sensor system with a sampling rate of 4000 Hz. The elbow joint angle was calculated using a marker- based 2D Noraxon Ninox camera system with a sampling rate of 125 Hz. 

\begin{figure}[htp]
     \centering
     \begin{subfigure}{0.3\columnwidth}
         \centering
         {\includegraphics[scale=0.2]{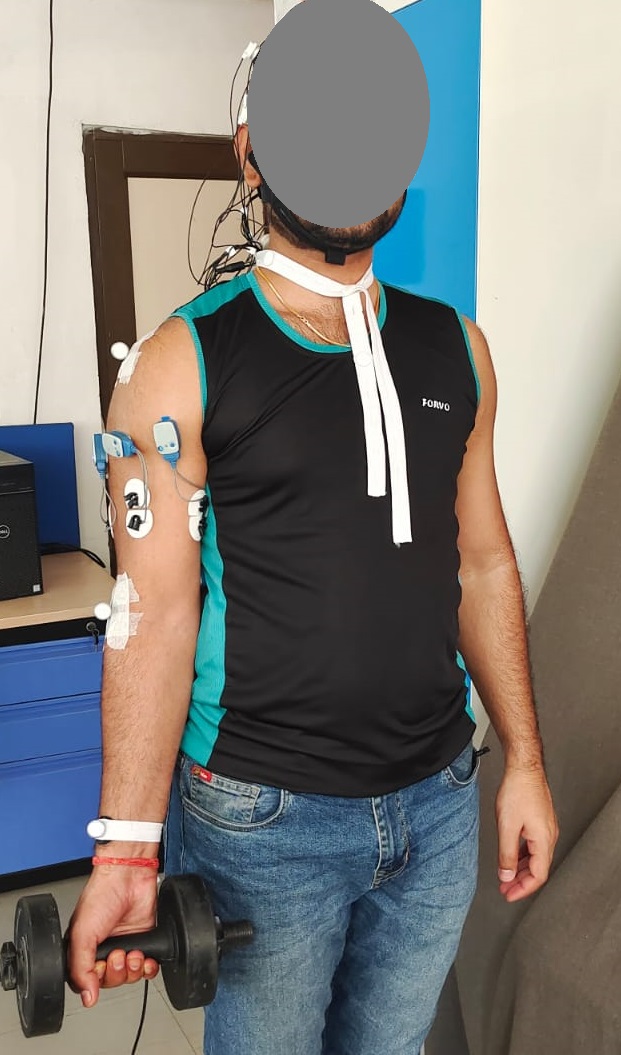}}\quad
         \caption{}
         \label{fig: Ayush data recording pic}
     \end{subfigure}
     \begin{subfigure}{0.3\columnwidth}
         \centering
         {\includegraphics[scale=0.167]{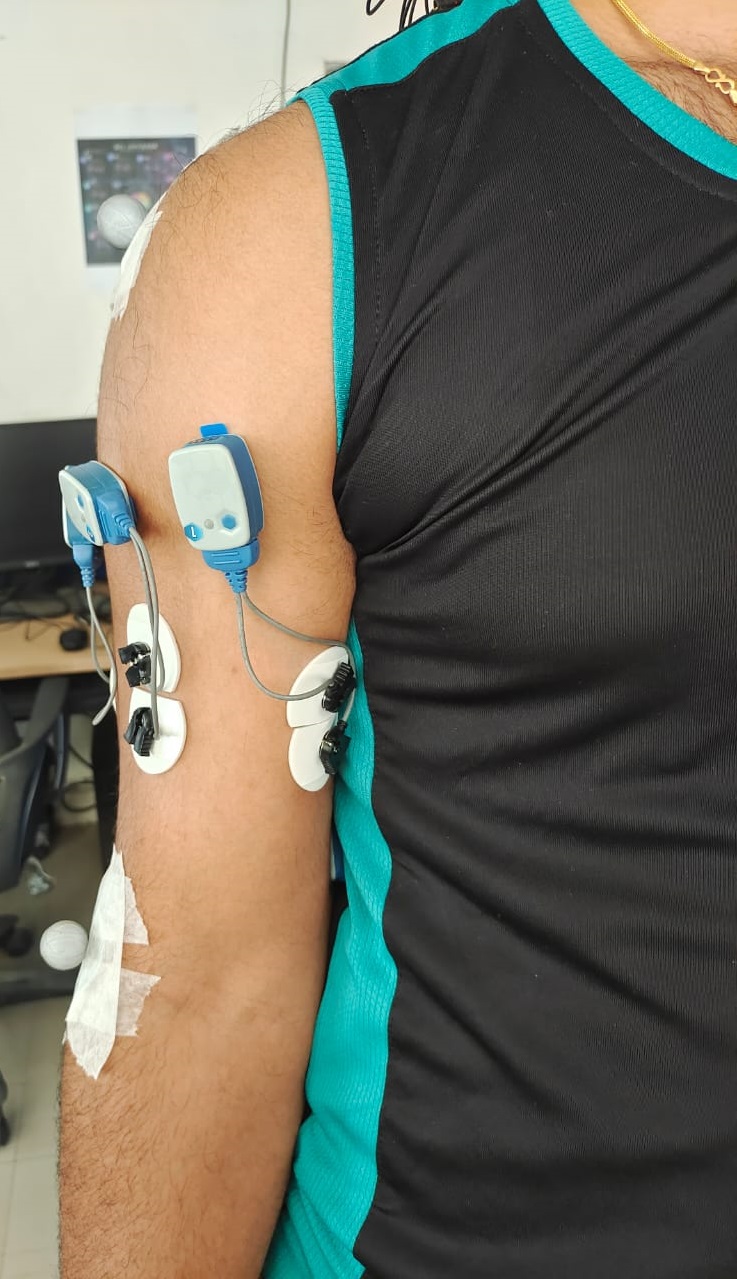}}\quad
         \caption{}
         \label{fig: EMG Placement front}
     \end{subfigure}
     \begin{subfigure}{0.3\columnwidth}
         \centering
         {\includegraphics[scale=0.18]{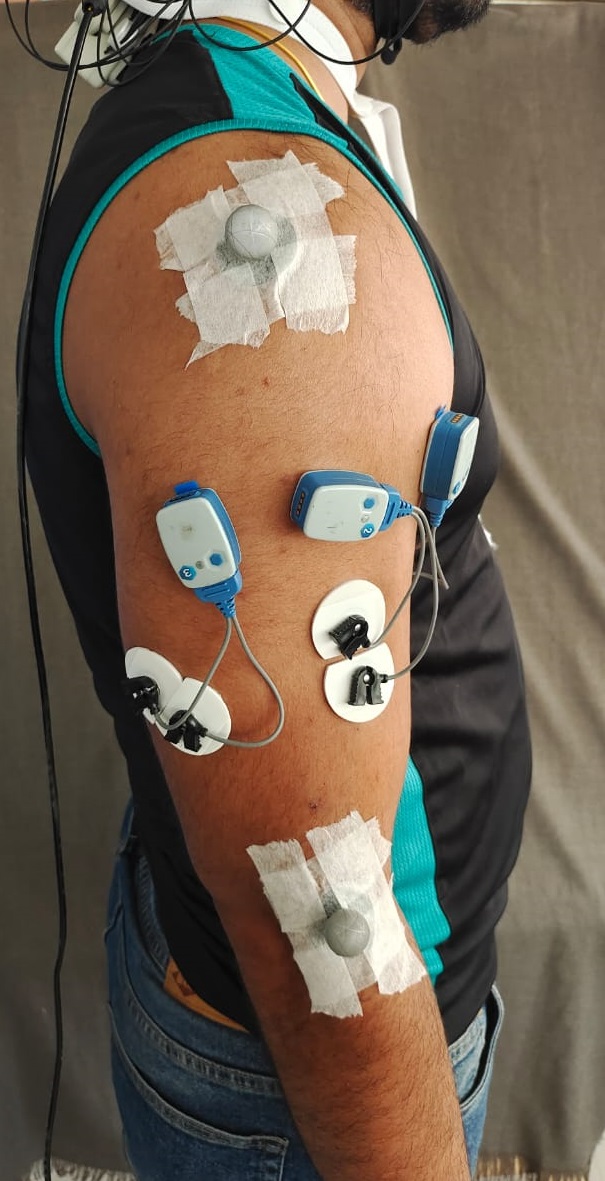}}
         \caption{}
         \label{fig: EMG Placement side}
     \end{subfigure}
      \caption{Experimental setup for EMG and joint angle data recording. (a) Four EMG electrodes and four IR markers were used on the hand. (b) and (c) Closer view of EMG electrode placement on biceps and triceps muscles. The task was elbow FE movement with 0, 2, and 4kg loads.}
        \label{fig:data recording setup}
\end{figure}

To record the EMG signals, four electrodes were placed, two on the biceps brachii muscle and two on the triceps brachii muscle, as shown in Fig. \ref{fig:data recording setup}. The placement of EMG electrodes and skin surface preparation followed the guidelines established in \cite{konrad2005abc}. Additionally, four reflective markers were placed to capture shoulder and elbow joint angles during the motion. 

The data recording paradigm consisted of an initial 10-15 second period where the participant stood in an upright posture. Each trial began with a cross appearing on the screen accompanied by a beep sound, signaling the start of the trial. A visual cue was then provided to prompt the participant to initiate the motion. The trial was manually ended when the subject completed the motion. A resting phase of 2 seconds followed the end of the hand motion. In total, 12 runs were recorded, with 4 runs for each load 0kg, 2kg, and 4kg, and each run included 10 trials.

The joint angle and electromyography (EMG) data recorded from the subject were specifically for the elbow FE movement. To enhance the quality of the joint angle data, a Gaussian filter with a standard deviation of 10 and a window size of 6 is applied. For the EMG data, a series of processing steps were implemented. Firstly, the raw EMGs underwent band-pass filtering within the frequency range of 10 to 450 Hz. Then, a full-wave rectification process is performed, followed by 7 Hz low-pass filtering \cite{sartori2012emg}. The outcome of these steps are linear envelopes that represented the amplitude of the EMG signals. To normalize these envelopes, they are divided by the maximum voluntary contraction (MVC) EMG values obtained from the set of MVC recorded trials. This normalization process helps standardize the EMG data across different trials and subjects.

\subsection {Benchmark joint angle and torque}

In this study, the benchmark joint angles are acquired by capturing the motion of the human arm as it performs elbow flexion-extension (FE) movements. Measuring multi-joint torque directly from the human body is challenging due to the impracticality of installing torque sensors inside the joints. Instead, the benchmark torque is determined using an inverse dynamics approach, which is elaborated on in Section II (C) and specifically referenced in Equation \ref{Equation of motion}. It's worth noting that the motion capture and inverse dynamics methods are widely recognized as the gold standard for this purpose, as acknowledged in previous studies \cite{lloyd2003emg, durandau2019voluntary}.

To estimate the mass and the center of mass for various body segments, a method described in \cite{herman2016physics} is employed. This process entails both single and double differentiation of the joint angles through the use of a central difference scheme, ultimately providing angular velocity and acceleration values. The motion capture technology is used to record the elbow and shoulder joint angles while performing elbow FE motions under different loads, including 0kg, 2kg, and 4kg.

To calculate the mass matrix of the system, a set of regression equations is employed to determine the Moment of Inertia (MOI) around the transverse axis passing through the center of mass for each segment. These regressions rely on anthropometric measurements as predictors, encompassing parameters such as the participants' height, weight, acromion-radiale length, arm circumference, biceps circumference, forearm circumference, wrist circumference, wrist circumference, and radiale-styloid length, all as described in \cite{hinrichs1985regression}.

\subsection {Baseline ANN and PINN model training}
To evaluate the performance of the suggested physics-informed deep learning framework, a fully connected feed-forward neural network (ANN) is employed as the baseline for comparison. This comparison primarily examines the contrast between physics-uninformed and physics-informed models.

In this configuration, the ANN is composed of four hidden layers, each containing 75 neurons. It possesses an input layer with five features (four EMG and one time) as inputs and an output layer with two neurons dedicated to the prediction of elbow and shoulder joint angles. During the training of the ANN model, the loss function used is the mean square error (MSE), calculated as indicated in Equation \ref{Equation data loss}.

The training process of the ANN model involves employing the Adam optimizer with momentum and relies on mini-batches, with each batch consisting of 75 data samples. Training continues for a total of 1000 epochs, with a specified step size of 300 epochs. This step size means that the learning rate is adjusted every 300 epochs. The reduction factor, referred to as gamma, is set to 0.8, which means that the learning rate is multiplied by 0.8 at each adjustment. This strategy allows the model to start with a relatively high initial learning rate, facilitating rapid convergence in the early training stages. As training proceeds, the learning rate decreases, enabling the model to fine-tune its parameters and approach an optimal solution.

\begin{figure}[H]
  \begin{center}
  \includegraphics[width=3.5in]{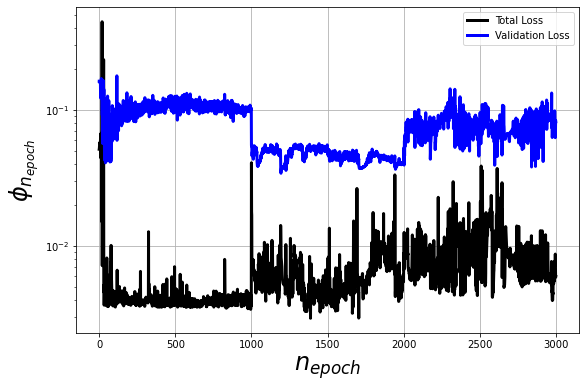}\\
  \caption{Representative of the loss during the training of the proposed physics-informed model. Each consecutive block of 1000 epochs corresponds to the training for a specific load condition. In total, 3000 epochs were utilized for training across the three different loading conditions.}\label{loss}
  \end{center}
\end{figure}

For the PINN model training, the neural network architecture and hyperparameters remain the same as the ANN. However, during the network's training, a customized loss function is utilized, as specified in Equation \ref{equation custom loss}. The convergence of the total loss of the proposed framework is demonstrated in Fig.\ref{loss}. The overall dataset comprises 12 runs, with 4 runs for each load condition (0kg, 2kg, and 4kg), and each run includes 10 trials. Out of the total dataset, 2 runs are used for training, while one run each is allocated for validation and testing purposes, respectively.

\subsection {Joint angle prediction and Evaluation criteria}
Once the models are trained, the test dataset is utilized to forecast joint angles using both the ANN and PINN models. To assess the predictive performance of the proposed framework, the root mean square error (RMSE) is initially utilized as the metric. To be more precise, RMSE gauges the disparities in magnitude between the predicted values and the reference or benchmark values.
\begin{equation}
RMSE = \sqrt{\frac{1}{n}\sum_{i=1}^{n}(y_i - \hat{y}_i)^2}
\end{equation}
Here, $n$ is the number of data points, $y_i$ represents the actual observed values, $\hat{y}_i$ represents the predicted values.

Pearson's correlation coefficient ($R$) is also utilized as an additional metric.
\begin{equation}
R = \frac{\sum_{i=1}^{n}(y_i - \bar{y})(\hat{y}_i - \bar{\hat{y}_i})}{\sqrt{\sum_{i=1}^{n}(y_i - \bar{y})^2}\sqrt{\sum_{i=1}^{n}(\hat{y}_i - \bar{\hat{y}_i})^2}}
\end{equation}

Here, $R$ represents Pearson's correlation coefficient, $y_i$ and $\hat{y}_i$ are the data values.
$\bar{y_i}$ and $\bar{\hat{y}_i}$ are the means of the respective data sets.

\section{RESULTS}
In this section, we assess the effectiveness of the proposed framework in concurrently predicting the angles of both the elbow and shoulder joints under varying loading conditions. We compare the proposed framework to a baseline method that utilizes feed-forward Artificial Neural Networks (ANN). Our evaluation commences with an overview of the training process for the proposed framework. Following that, we conduct a comprehensive comparison to present the predicted results from both the proposed framework and the baseline method. This includes presenting results for the predicted joint angles and providing detailed and average predictions for four subjects.

To illustrate the convergence pattern of the proposed framework, we display the progression of both total and validation loss for all joints and loads during the training phase, as shown in Figure \ref{loss}. When predicting both shoulder and elbow joint angles simultaneously under 0kg, 2kg, and 4kg loads, the total training and validation loss reaches a very low value after approximately 1000 iterations for each, with minor local oscillations occurring as training progresses. This behavior is largely influenced by our choice of a batch size of 75 during Physics-Informed Neural Network (PINN) training. While this batch size aids in better learning the data distribution, it can also lead to these minor oscillations. It's noteworthy that similar convergence patterns have also been observed in the case of the Artificial Neural Network (ANN).

\begin{figure}[H]
     \centering
     \begin{subfigure}[b]{0.497\columnwidth}
         \centering
         {\includegraphics[trim=0.1cm 6.5cm 0.1cm 6.6cm, clip=true,width=1.1\linewidth]{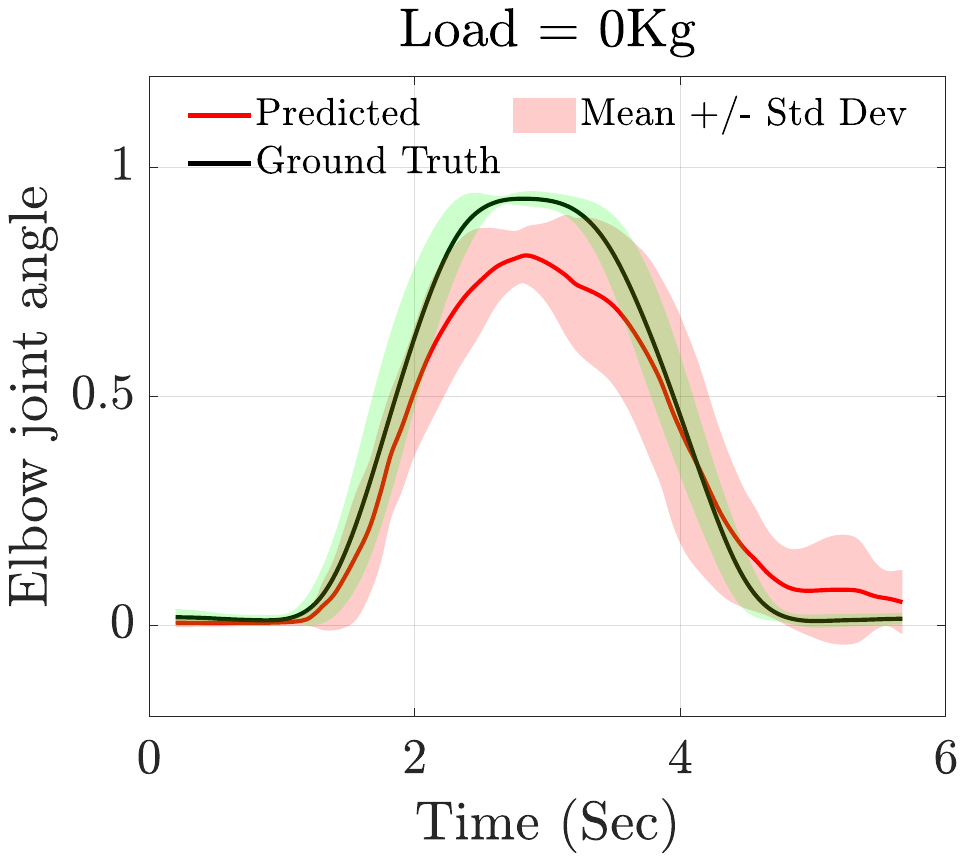}}
         \caption{Elbow joint angle}
         \label{figure:0kg_elbow}
     \end{subfigure}
     \hfill
     \begin{subfigure}[b]{0.483\columnwidth}
         \centering
         {\includegraphics[trim=0.1cm 6.5cm 0.1cm 6.5cm, clip=true,width=1.1\linewidth]{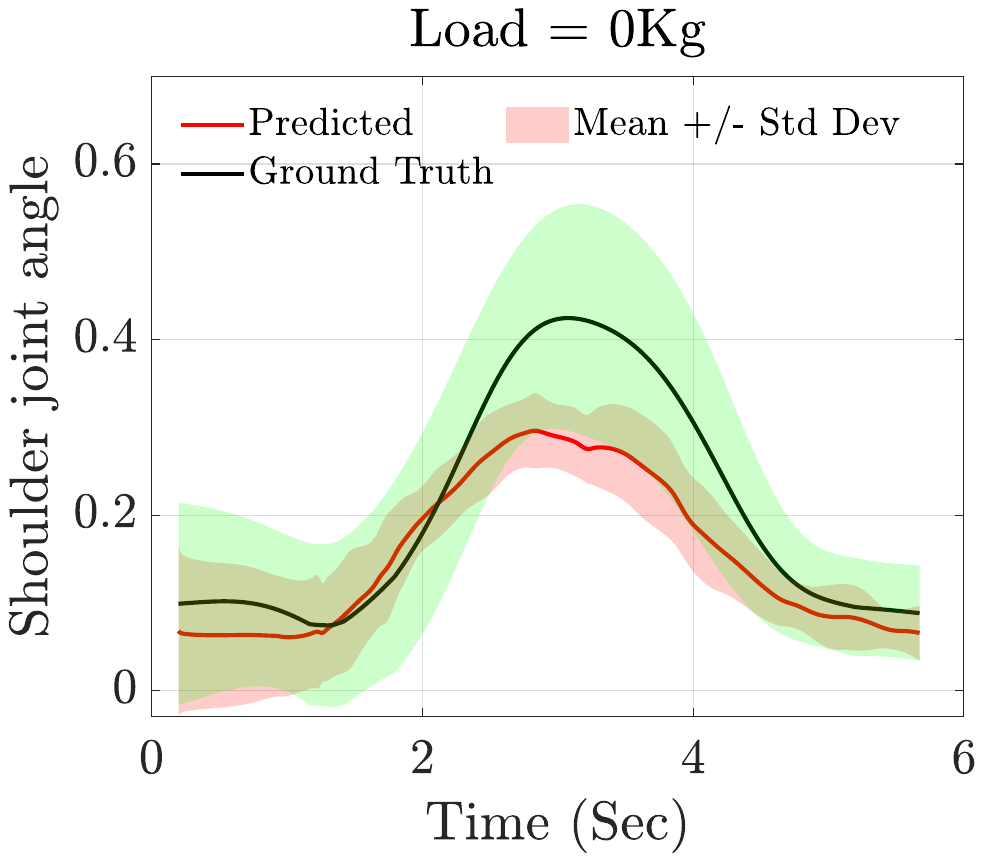}}
         \caption{Shoulder joint angle}
         \label{figure:0kg_shoulder}
     \end{subfigure}
     \centering
     \begin{subfigure}[b]{0.49\columnwidth}
         \centering
         {\includegraphics[trim=0.1cm 6.5cm 0.1cm 6.5cm, clip=true,width=1.1\linewidth]{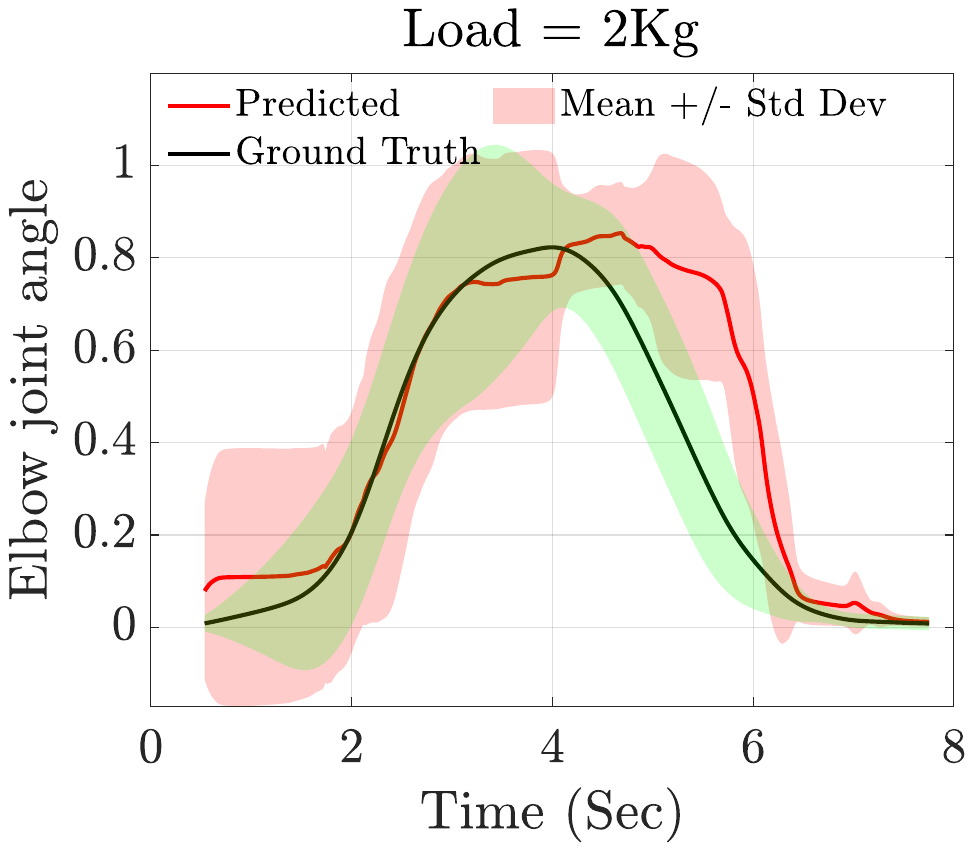}}
         \caption{Elbow joint angle}
         \label{figure: 2kg_elbow}
     \end{subfigure}
     \hfill
     \begin{subfigure}[b]{0.49\columnwidth}
         \centering
         {\includegraphics[trim=0.1cm 6.5cm 0.1cm 6.5cm, clip=true,width=1.1\linewidth]{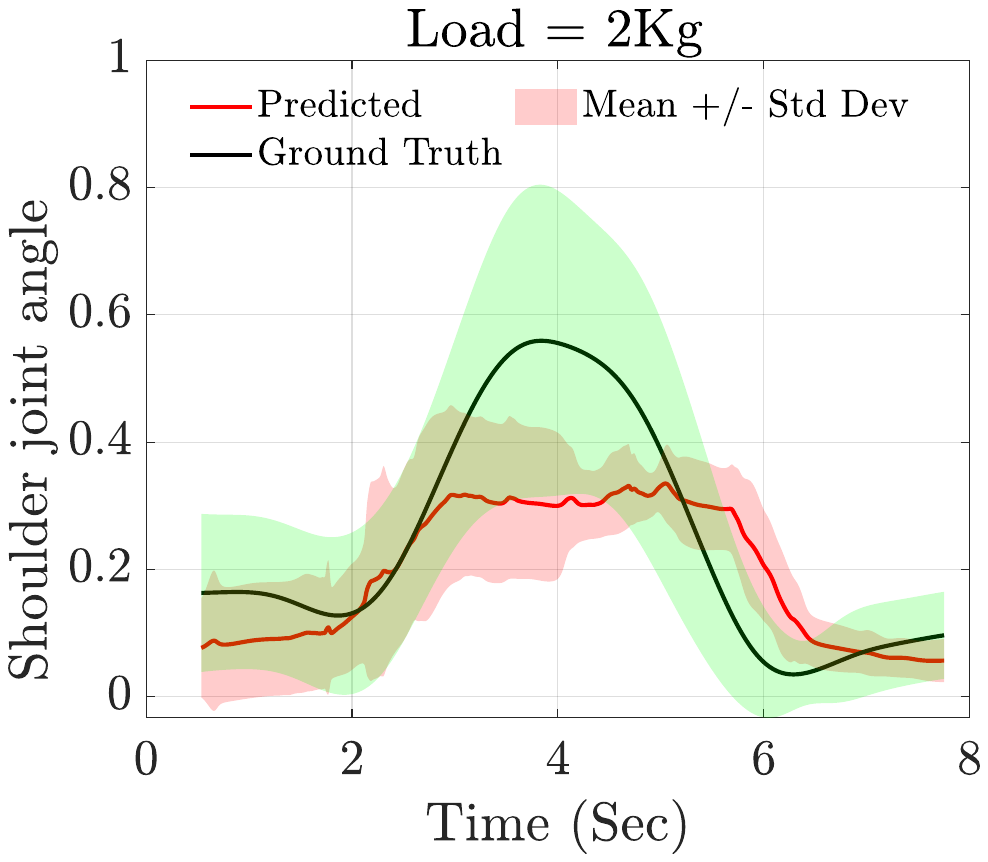}}
         \caption{Shoulder joint angle}
         \label{figure:2kg_shoulder}
     \end{subfigure}
     \centering
     \begin{subfigure}[b]{0.49\columnwidth}
         \centering
         {\includegraphics[trim=0.1cm 6.5cm 0.1cm 6.5cm, clip=true,width=1.1\linewidth]{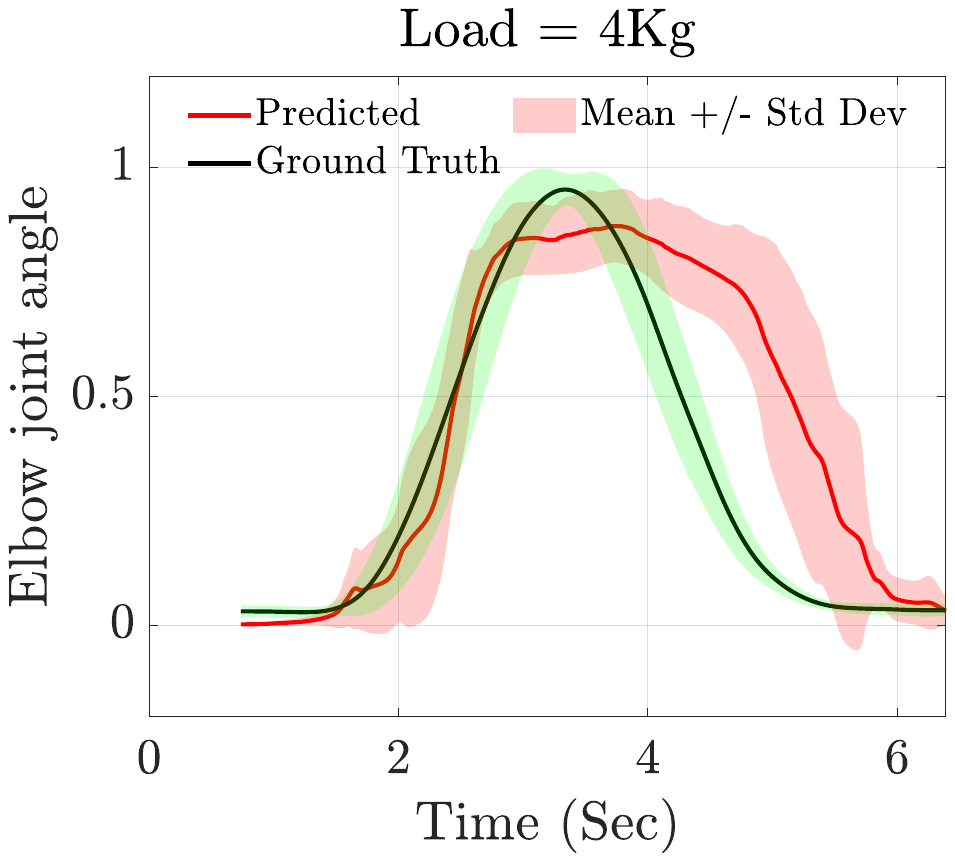}}
         \caption{Elbow joint angle}
         \label{figure:4kg_elbow}
     \end{subfigure}
     \hfill
     \begin{subfigure}[b]{0.49\columnwidth}
         \centering
         {\includegraphics[trim=0.1cm 6.5cm 0.1cm 6.5cm, clip=true,width=1.1\linewidth]{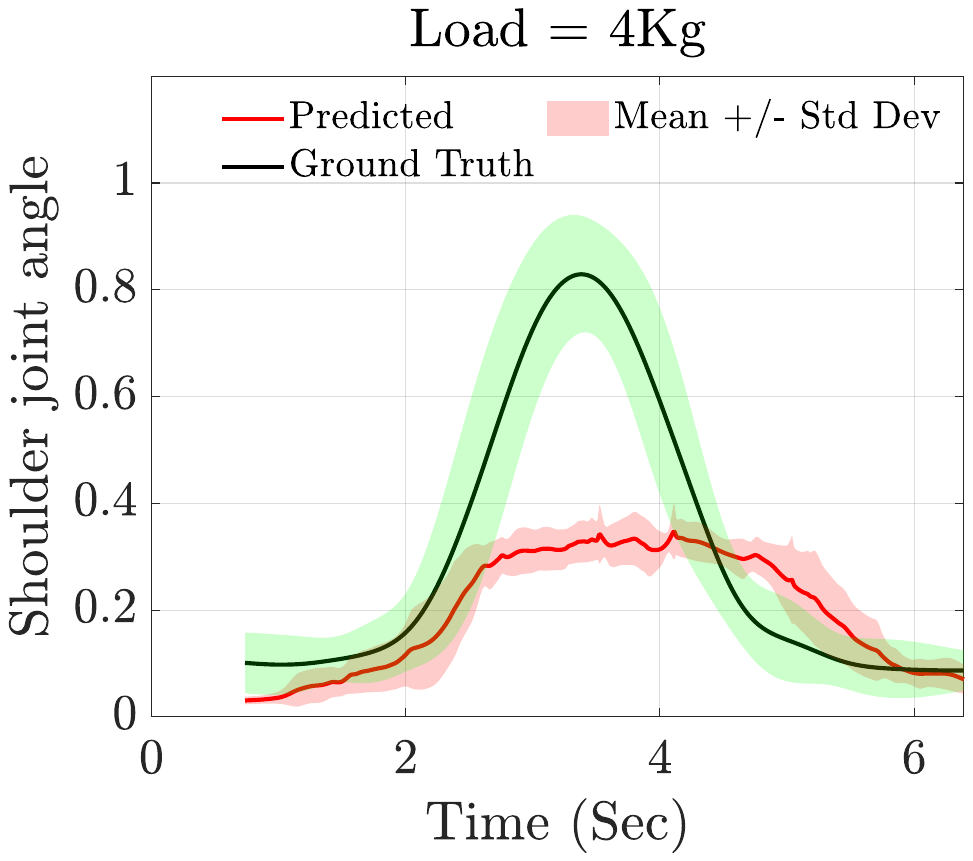}}
         \caption{Shoulder joint angle}
         \label{figure: 4kg_shoulder}
     \end{subfigure}
      \caption{Comparing PINN joint angle predictions (Normalized with respect to 4kg load joint angle) for subject 1 (S1), the red line signifies model prediction with shaded uncertainty bounds. The black line denotes angle sourced from motion capture (ground truth). Here, Fig. \ref{figure:0kg_elbow}\&\ref{figure:0kg_shoulder}, \ref{figure: 2kg_elbow}\&\ref{figure:2kg_shoulder} and \ref{figure:4kg_elbow}\&\ref{figure: 4kg_shoulder} is for $No$-load, $2kg$-load and $4kg$-load respectively for a single repetition of elbow flexion-extension (FE).}
        \label{fig:Result PINN}
\end{figure}

We have carried out thorough comparisons between the proposed framework and the baseline (ANN) method. In Figure \ref{fig:Result PINN}, we present representative results for one subject (S1) obtained from the proposed framework for both the shoulder joint and elbow joint under 0kg, 2kg, and 4kg loading conditions, which are then compared with the corresponding ground truth. These results have been derived from the testing dataset, which comprises EMG signals from muscles such as the Biceps long head, Biceps short head, Triceps long head, and Triceps lateral head, covering both joints. Upon a meticulous analysis of Figure \ref{fig:Result PINN}, it becomes evident that the predicted joint angle values closely align with the ground truth data. This observation underscores the remarkable dynamic tracking capability of the proposed framework.

\begin{figure}[h!tb]
	\centering
	\begin{subfigure}[b]{0.27\linewidth}
		\includegraphics[trim=0.1cm 0.1cm 0.1cm 0.1cm, clip=true,width=1\textwidth]{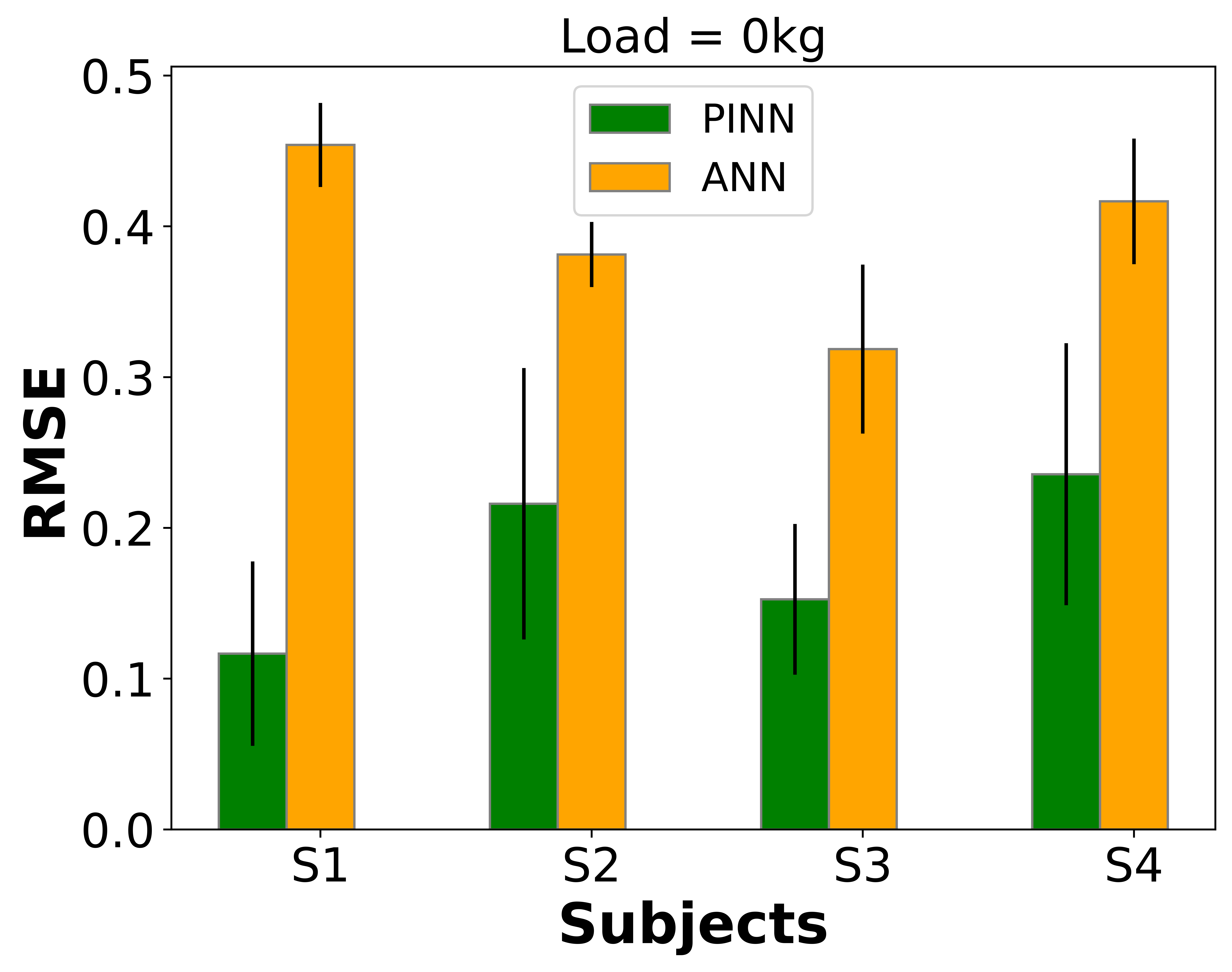}
		\caption{No Load}
		\label{fig:subfigA}
	\end{subfigure}
	\begin{subfigure}[b]{0.27\linewidth}
		\includegraphics[trim=0.2cm 0.1cm 0.1cm 0.1cm, clip=true,width=1\linewidth]{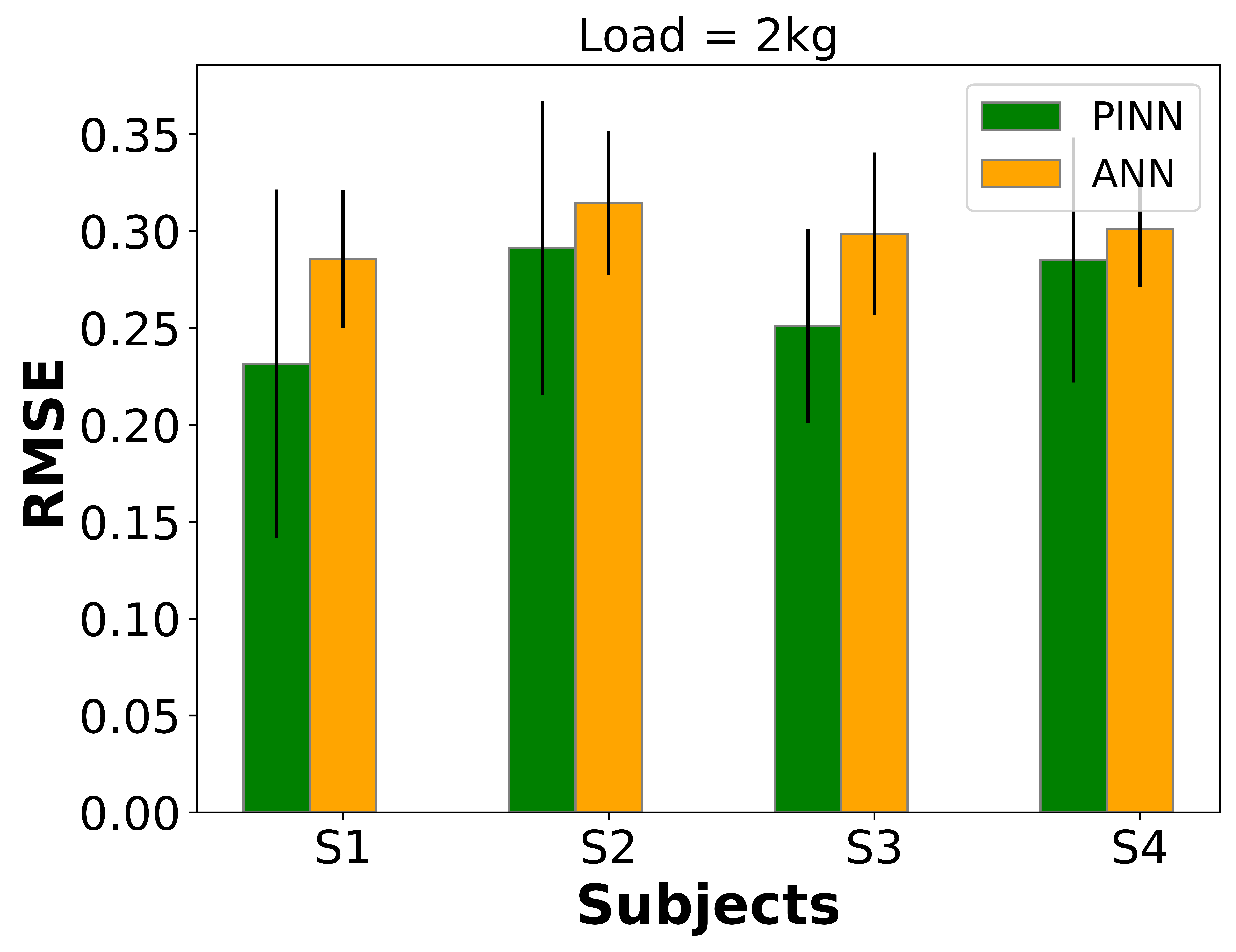}
		\caption{2kg Load}
		\label{fig:subfigB}
	\end{subfigure}
	\begin{subfigure}[b]{0.27\linewidth}
	        \includegraphics[trim=0.2cm 0.1cm 0.1cm 0.1cm, clip=true,width=1\linewidth]{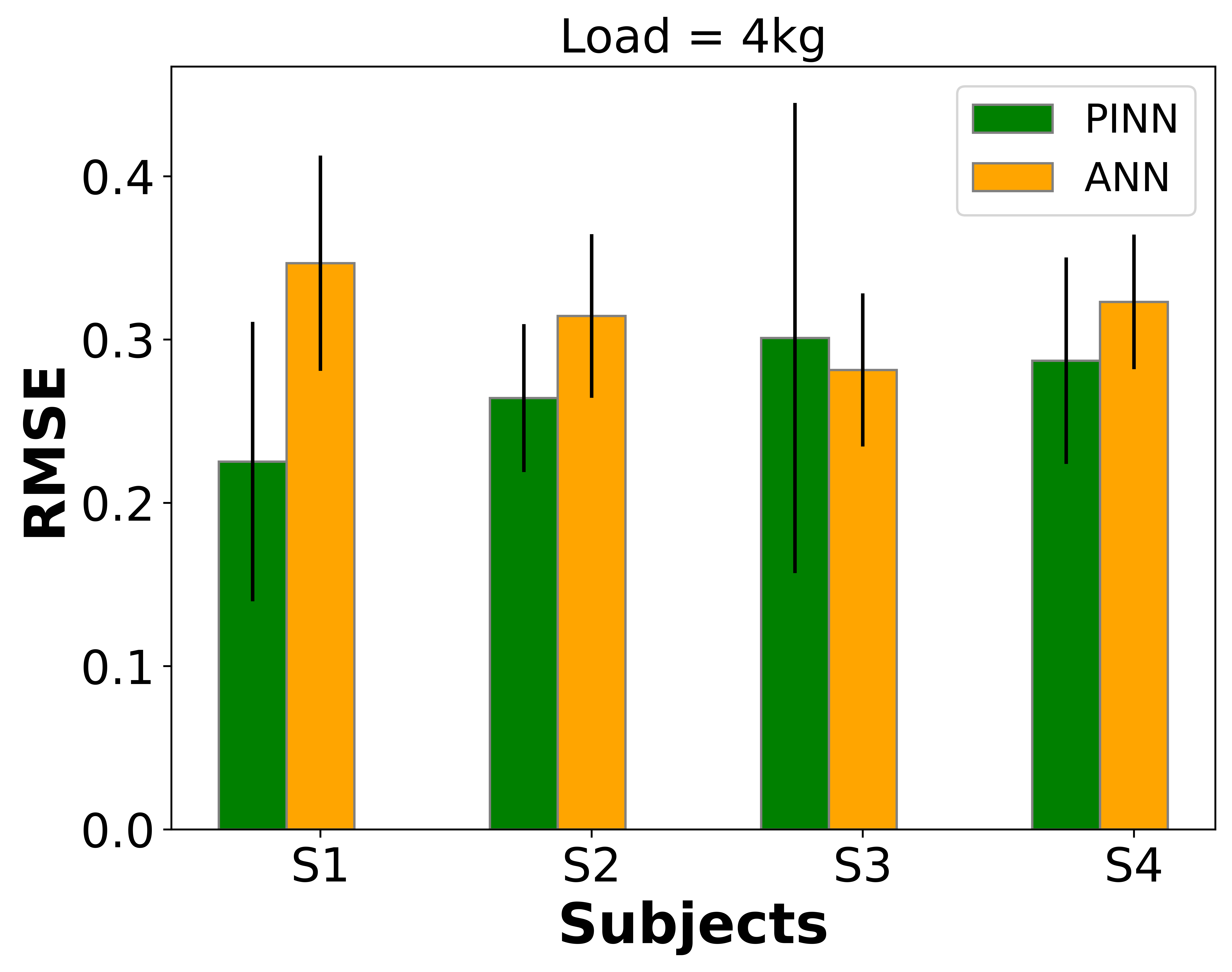}
	        \caption{4Kg Load}
	        \label{fig:subfigC}
         \end{subfigure}
	\begin{subfigure}[b]{0.27\linewidth}
		\includegraphics[trim=0.11cm 0.1cm 0.1cm 0.1cm, clip=true,width=1\linewidth]{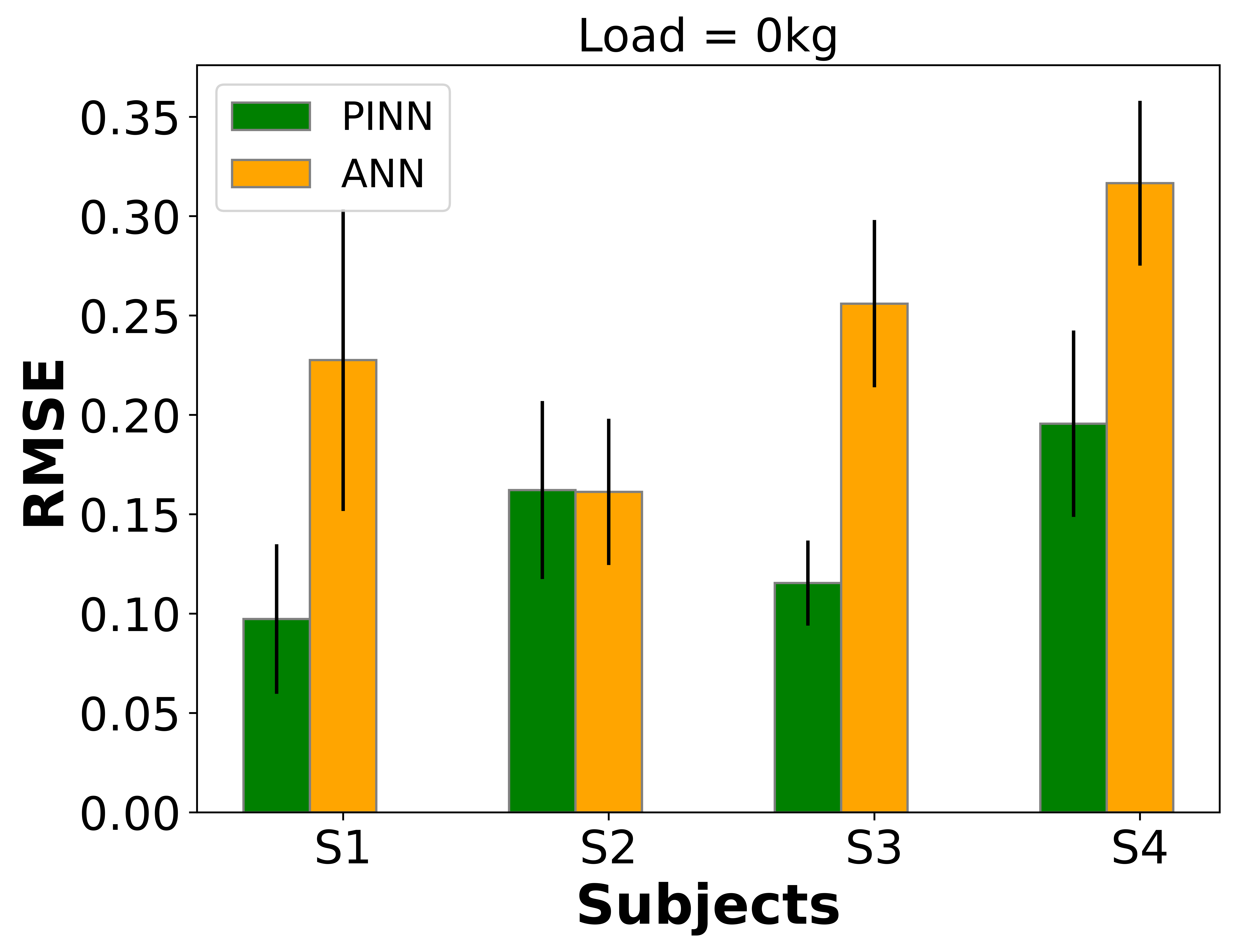}
		\caption{No Load}
		\label{fig:subfigD}
	\end{subfigure}
	\begin{subfigure}[b]{0.27\linewidth}
		\includegraphics[trim=0.11cm 0.1cm 0.1cm 0.1cm, clip=true,width=1\linewidth]{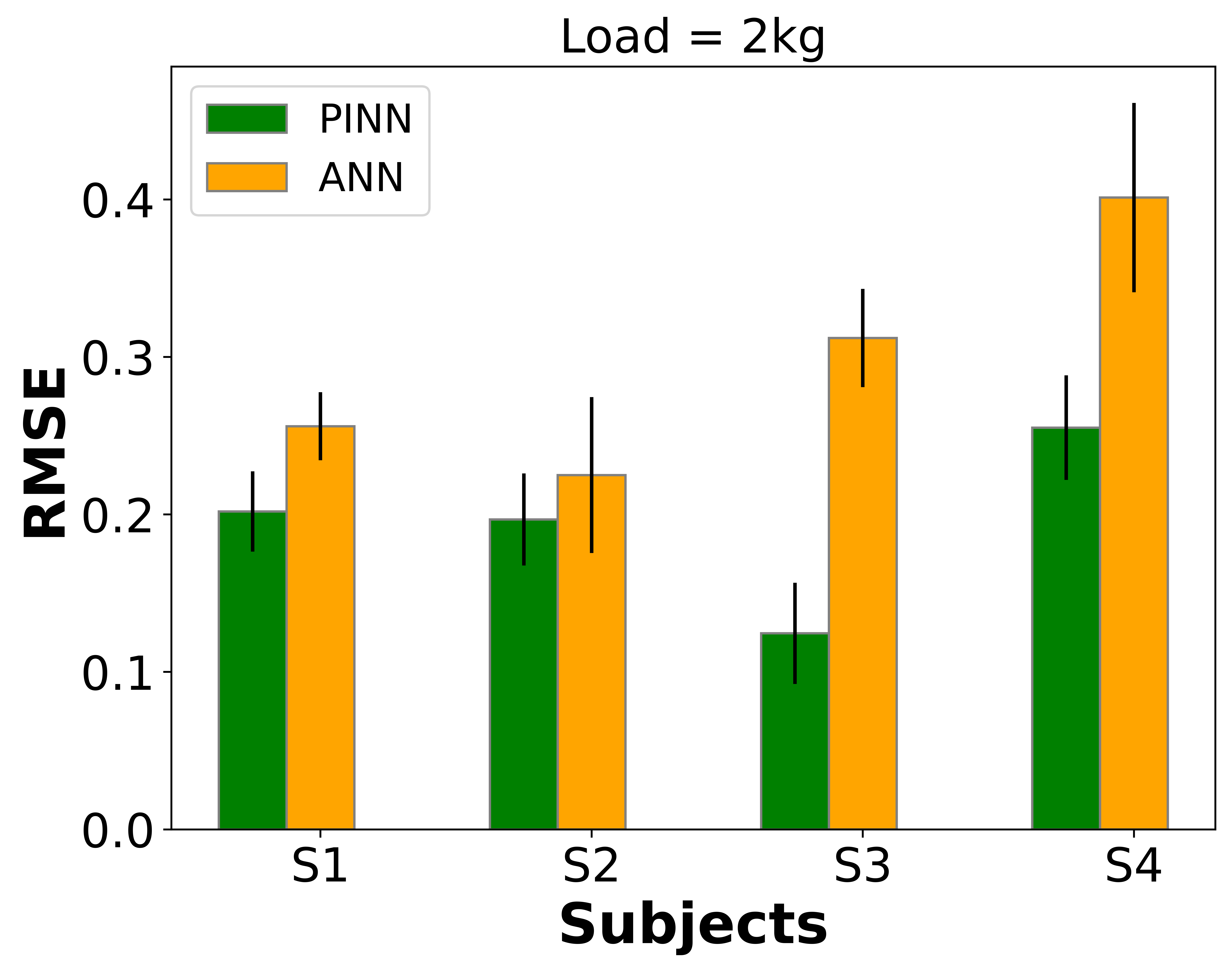}
		\caption{2kg Load}
		\label{fig:subfigE}
	\end{subfigure}
	\begin{subfigure}[b]{0.27\linewidth}
	        \includegraphics[trim=0.11cm 0.1cm 0.1cm 0.1cm, clip=true,width=1\linewidth]{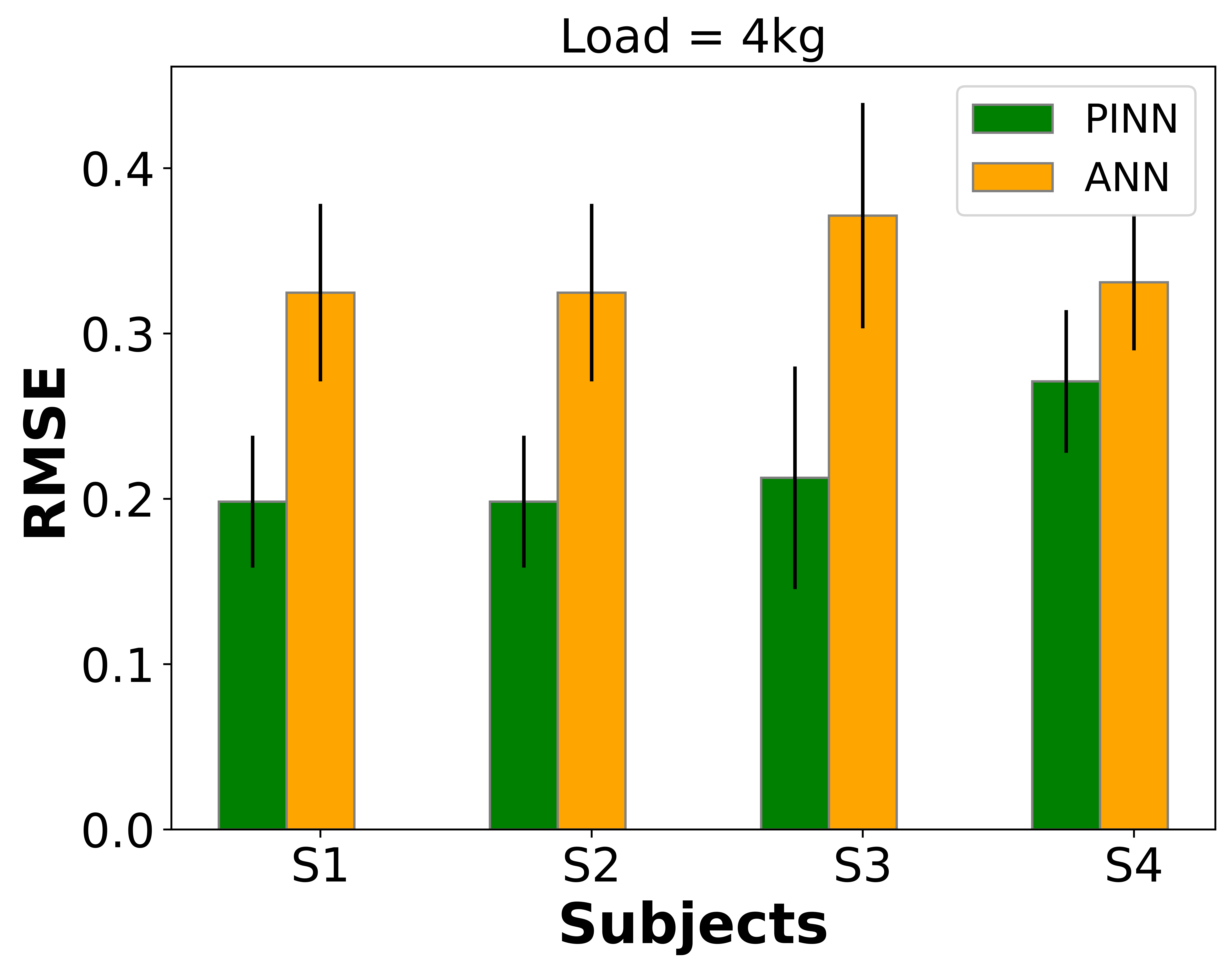}
	        \caption{4Kg Load}
	        \label{fig:subfigF}
         \end{subfigure}
	\caption{Histograms in Figures (Fig. \ref{fig:subfigA}-\ref{fig:subfigC}) and (Fig. \ref{fig:subfigD}-\ref{fig:subfigF}) depict the Root mean square errors (RMSEs) across all four subjects (S1-S4) for the shoulder and elbow joint angles, respectively. These histograms illustrate differences between normalized predicted and ground truth joint angles under three loads (No load, 2kg, 4kg), using the PINN model (green bars) and the ANN model (orange).}
	\label{fig:RMSE BARS}
\end{figure}

To further evaluate the performance of the proposed framework, we provide comprehensive comparisons for all subjects between the proposed framework and the baseline method in Table \ref{Table: CORR COEFF} and Figure \ref{fig:RMSE BARS}. In most instances, the proposed framework produces smaller Root Mean Square Errors (RMSEs) and higher Pearson correlation coefficients, underscoring the robustness of this approach.

Compared to fully connected feed-forward network-based method, the proposed framework excels by achieving the best predictive performance. This success is attributed to the incorporation of physics laws, which facilitates the generalization of the model for multiple loading conditions by penalizing the Artificial Neural Network (ANN) within the proposed framework. As a result, the performance is not solely reliant on the traditional Mean Square Error (MSE) loss but can also be improved through the incorporation of the physics-based loss.

The importance of the proposed framework is evident from the outcomes of the statistical analysis provided in Table \ref{Table: CORR COEFF} and Figure \ref{fig:RMSE BARS}.
Figure \ref{fig:RMSE BARS} illustrates the mean Root Mean Square Errors (RMSEs) for the shoulder and elbow joints in both the proposed framework and the baseline method, encompassing all participants. As evident in Figure \ref{fig:RMSE BARS}, the proposed framework demonstrates impressive performance, distinguished by decreased standard deviations and consistently predictable outcomes with minimal fluctuations.

Remarkably, despite having a simpler neural network architecture, the proposed framework achieves comparable performance when compared to a feed-forward Artificial Neural Network (ANN). This achievement is credited to the integration of the fundamental physical relationships between the predicted variables within the data-driven model.

\begin{table*}[]
\centering
\begin{tabular}{ccccccc}
\hline
\multicolumn{7}{c}{\textbf{Coefficient of correlation (R)}}                                                          \\ \hline
Joint     & \multicolumn{6}{c}{\textbf{Shoulder}}                                                                    \\ \hline
Load (Kg) & \multicolumn{2}{c}{0}           & \multicolumn{2}{c}{2}         & \multicolumn{2}{c}{4}         \\ \hline
Model     & \textbf{PINN}           & ANN            & \textbf{PINN}          & ANN           & \textbf{PINN}          & ANN           \\ \hline
S1        & 0.90 $\pm$0.08 & 0.37$\pm$0.48  & 0.68$\pm$0.14 & 0.35$\pm$0.35 & 0.76$\pm$0.10 & 0.25$\pm$0.40 \\
S2        & 0.78$\pm$0.15  & 0.48$\pm$0.13  & 0.65$\pm$0.08 & 0.27$\pm$0.29 & 0.71$\pm$0.09 & 0.38$\pm$0.21 \\
S3        & 0.71$\pm$0.11  & 0.40$\pm$0.22  & 0.56$\pm$0.27 & 0.32$\pm$0.15 & 0.58$\pm$0.21 & 0.42$\pm$0.19 \\ 
S4        & 0.58$\pm$0.36  & 0.15$\pm$0.32  & 0.41$\pm$0.19 & 0.17$\pm$0.29 & 0.46$\pm$0.34 & 0.15$\pm$0.30 \\ \hline
\textbf{Mean}      & \textbf{0.74 $\pm$0.17}                & 0.35 $\pm$0.28                & \textbf{0.58 $\pm$0.17}                & 0.28 $\pm$0.27               & \textbf{0.63 $\pm$0.18}              & 0.30 $\pm$0.27              \\ \hline \\
Joint     & \multicolumn{6}{c}{\textbf{Elbow}}                                                                       \\ \hline
Load (Kg) & \multicolumn{2}{c}{0}           & \multicolumn{2}{c}{2}         & \multicolumn{2}{c}{4}         \\ \hline
Model     & \textbf{PINN}           & ANN            & \textbf{PINN}          & ANN           & \textbf{PINN}          & ANN           \\ \hline
S1        & 0.95$\pm$0.05  & 0.67$\pm$0.07  & 0.83$\pm$0.15 & 0.42$\pm$0.16 & 0.83$\pm$0.08 & 0.35$\pm$0.29 \\
S2        & 0.86$\pm$0.13  & -0.09$\pm$0.19 & 0.78$\pm$0.08 & 0.38$\pm$0.15 & 0.75$\pm$0.02 & 0.17$\pm$0.19 \\
S3        & 0.74$\pm$0.20  & 0.15$\pm$0.17  & 0.72$\pm$0.13 & 0.23$\pm$0.25 & 0.80$\pm$0.08 & 0.07$\pm$0.35 \\
S4        & 0.76$\pm$0.11  & 0.27$\pm$0.15  & 0.68$\pm$0.16 & 0.37$\pm$0.12 & 0.73$\pm$0.11 & 0.13$\pm$0.26 \\ \hline
\textbf{Mean}      & \textbf{0.83 $\pm$0.12}               &  0.25 $\pm$0.14              &  \textbf{0.72 $\pm$0.13}             & 0.35 $\pm$0.17              & \textbf{0.78 $\pm$0.07}              & 0.18 $\pm$0.27              \\ \hline
\end{tabular}
\caption{{Summary of the correlation coefficient ($R$): The prediction of shoulder and elbow joint angles across all four subjects (S1 - S4) under all three loading conditions (0 kg, 2 kg, and 4 kg). A comparative analysis of the correlation coefficient has been conducted for both Physics-Informed Neural Networks (PINN) and Artificial Neural Networks (ANN).}}\label{Table: CORR COEFF}
\end{table*}

This accomplishment offers inspiration for future efforts, implying that enhancements in performance could be achieved by incorporating a more extensive volume of EMG data from a broader array of muscles spanning across the joints, even delving into the inclusion of deeper muscles, within the training dataset. Additionally, exploring and comparing deeper neural network architectures can lead to a more profound integration of musculoskeletal model knowledge into neural networks.
\section{Discussion}
The central objective of this research is to predict the kinematics of multiple joints involved in the dynamic process of elbow flexion-extension during upper limb motion employing physics-informed neural network modeling. In this pursuit, we harness the rich dataset of electromyography (EMG) signals that spans across these joints, aiming to discern the intricate relationships and predictive potential of this modeling paradigm. While the literature has previously witnessed promising applications of similar modeling techniques in estimating muscle forces and joint kinematics, as elaborated upon in prior studies \cite{zhang2022physics,taneja2022feature}, it is crucial to note that these prior methodologies are often confined to single-joint scenarios and do not account for the complexity introduced by dynamic movement conditions.

In the context of our investigation, we have effectively overcome this constraint by employing a Physics-Informed Neural Network (PINN) to concurrently predict the angles of both the shoulder and elbow joints. Importantly, we have done so across a spectrum of diverse loading conditions, as illustrated in Figure \ref{fig:Result PINN}. The outcomes of our study substantiate the primary hypothesis, manifesting that PINN is adept at precise estimation of multi-joint angles pertaining to upper limb motion, even in the face of varying loading conditions and relying on muscle electromyography (EMG) signals. This method, therefore, represents a comprehensive and promising approach to the prediction of joint angles through the utilization of EMG signals. 

It's essential to emphasize that while our approach offers a level of generalization for a specific subject, the need for individualized training should not be underestimated. This underscores the core principle of our proposed framework, which involves gathering subject-specific data and parameters. In essence, our framework assimilates and understands the complexities of the musculoskeletal system, as described in Equation \ref{Equation of motion}, while also considering the distinct anthropometric features of each individual. In doing so, it effectively accounts for the intricacies of the anatomy and physiology inherent to each subject's system.

Moreover, the results of our investigation undeniably demonstrate the remarkable capability of the Physics-Informed Neural Network (PINN) model in learning the intricacies of musculoskeletal forward dynamics, all of which are guided by the information embedded in electromyography (EMG) signals. This proficiency is particularly evident in its ability to predict joint kinematics. However, it's important to note that the empirical evidence we have gathered from this study reveals an interesting observation. Specifically, our findings highlight that, within the framework of the current setup, the model's predictions regarding shoulder angles exhibit a slightly lower level of accuracy when compared to the precision achieved in estimating elbow joint angles.

The discernible difference in predictive accuracy can be attributed to a notable limitation inherent in our study. This limitation pertains to the unavailability of electromyography (EMG) signals emanating from deeper musculature, particularly the likes of the brachialis and triceps medial head. Given that these deeper musculature elements potentially exert an influence on the dynamics of the joint under scrutiny, their omission from our dataset resulted in an absence of pertinent EMG signals, thus contributing to the observed discrepancy in prediction accuracy.

Furthermore, it's important to underscore that when we delve into the analysis of shoulder joint movements and their accompanying EMG signals, the consideration of major shoulder flexor-extensor muscles assumes paramount significance. These muscles, including the deltoid, coracobrachialis, pectoralis major, latissimus dorsi, teres major, and teres minor, wield substantial impact on the dynamics of the shoulder joint. Their involvement and the nuances of their EMG signals are integral factors that can account for the variations in prediction accuracy in the context of shoulder joint kinematics.

In the process of training the Physics-Informed Neural Network (PINN) model, we adopted a distinctive approach for organizing the electromyography (EMG) signals. Specifically, we structured the data sequentially based on the load conditions, thereby maintaining a distinct separation between data points corresponding to different loads. Importantly, we refrained from intermixing the data. Furthermore, it is worth noting that the size of the training dataset remained constant throughout the training process. This methodology was chosen, guided by insights from prior experimental investigations \cite{zhang2022physics}, which indicated that our proposed framework exhibited a remarkable resilience to variations in training dataset size.

The key innovation in our methodology lay in the incorporation of physics-based domain knowledge as a penalization or regularization term within the loss function of the Artificial Neural Network (ANN). This strategic integration had the dual advantage of expediting the convergence of the network during training and concurrently diminishing the stringency of data requirements for achieving a specific level of performance. By blending the power of physics-based constraints with the adaptability of neural networks, our approach not only demonstrated faster convergence but also showcased its capacity to excel with a consistent dataset size.

\section{Conclusion}\label{section-4}
In summary, this paper introduces a novel approach, a data-driven framework infused with physics. This framework effectively integrates domain-specific knowledge of physics into the data-driven model, with a particular emphasis on modeling musculoskeletal forward dynamics. More specifically, we use this physics-based domain knowledge as flexible constraints to modify or regulate the loss function of the neural network. This integration enhances the model's resilience and ability to generalize, while also considerably reducing the computational resources required for model development.

Our extensive experiments, conducted on four distinct subjects of data for predicting joint angles, demonstrate the viability of this proposed framework. We believe that this framework represents a versatile methodology applicable not only to joint angle prediction but also to various other aspects within the musculoskeletal modeling field. Its potential extends to bridging the gap between laboratory prototypes and practical clinical applications.

\section*{Acknowledgements}
The authors would like to thank Anant Jain from the Indian Institute of Technology (IIT) Delhi for helping in data collection. This study is partly supported by the Joint Advanced Technology Centre (JATC)  at IIT Delhi (Grant no.: RP03830G), sponsored by the Ministry of Education (MoE), Govt. of India.

\section*{Declaration of competing interest}
The author(s) declared no potential conflicts of interest with
respect to the research, authorship, and/or publication of this
article.



\begin{thebibliography}{1}

\bibitem{farina2017man}
Dario Farina, Ivan Vujaklija, Massimo Sartori, Tam{\'a}s Kapelner, Francesco Negro, Ning Jiang, Konstantin Bergmeister, Arash Andalib, Jose Principe, and Oskar~C Aszmann.
\newblock Man/machine interface based on the discharge timings of spinal motor neurons after targeted muscle reinnervation.
\newblock {\em Nature biomedical engineering}, 1(2):0025, 2017.

\bibitem{koller2015learning}
Jeffrey~R Koller, Daniel~A Jacobs, Daniel~P Ferris, and C~David Remy.
\newblock Learning to walk with an adaptive gain proportional myoelectric controller for a robotic ankle exoskeleton.
\newblock {\em Journal of neuroengineering and rehabilitation}, 12(1):1--14, 2015.

\bibitem{muceli2011simultaneous}
Silvia Muceli and Dario Farina.
\newblock Simultaneous and proportional estimation of hand kinematics from emg during mirrored movements at multiple degrees-of-freedom.
\newblock {\em IEEE transactions on neural systems and rehabilitation engineering}, 20(3):371--378, 2011.

\bibitem{jung2021intramuscular}
Moon~Ki Jung, Silvia Muceli, Camila Rodrigues, {\'A}lvaro Meg{\'\i}a-Garc{\'\i}a, Alejandro Pascual-Valdunciel, Antonio~J Del-Ama, Angel Gil-Agudo, Juan~C Moreno, Filipe~Oliveira Barroso, Jos{\'e}~L Pons, et~al.
\newblock Intramuscular emg-driven musculoskeletal modelling: Towards implanted muscle interfacing in spinal cord injury patients.
\newblock {\em IEEE Transactions on Biomedical Engineering}, 69(1):63--74, 2021.

\bibitem{pau2012neuromuscular}
James~WL Pau, Shane~SQ Xie, and Andrew~J Pullan.
\newblock Neuromuscular interfacing: Establishing an emg-driven model for the human elbow joint.
\newblock {\em IEEE Transactions on biomedical engineering}, 59(9):2586--2593, 2012.

\bibitem{chadwick2008real}
Edward~K Chadwick, Dimitra Blana, Antonie~J van~den Bogert, Robert~F Kirsch, et~al.
\newblock A real-time, 3-d musculoskeletal model for dynamic simulation of arm movements.
\newblock {\em IEEE Transactions on Biomedical Engineering}, 56(4):941--948, 2008.

\bibitem{blana2017real}
Dimitra Blana, Edward~K Chadwick, Antonie~J van~den Bogert, and Wendy~M Murray.
\newblock Real-time simulation of hand motion for prosthesis control.
\newblock {\em Computer methods in biomechanics and biomedical engineering}, 20(5):540--549, 2017.

\bibitem{crouch2016lumped}
Dustin~L Crouch and He~Huang.
\newblock Lumped-parameter electromyogram-driven musculoskeletal hand model: A potential platform for real-time prosthesis control.
\newblock {\em Journal of biomechanics}, 49(16):3901--3907, 2016.

\bibitem{au2000emg}
Arthur~TC Au and Robert~F Kirsch.
\newblock Emg-based prediction of shoulder and elbow kinematics in able-bodied and spinal cord injured individuals.
\newblock {\em IEEE Transactions on rehabilitation engineering}, 8(4):471--480, 2000.

\bibitem{cote2019deep}
Ulysse C{\^o}t{\'e}-Allard, Cheikh~Latyr Fall, Alexandre Drouin, Alexandre Campeau-Lecours, Cl{\'e}ment Gosselin, Kyrre Glette, Fran{\c{c}}ois Laviolette, and Benoit Gosselin.
\newblock Deep learning for electromyographic hand gesture signal classification using transfer learning.
\newblock {\em IEEE transactions on neural systems and rehabilitation engineering}, 27(4):760--771, 2019.

\bibitem{ma2020continuous}
Chenfei Ma, Chuang Lin, Oluwarotimi~Williams Samuel, Lisheng Xu, and Guanglin Li.
\newblock Continuous estimation of upper limb joint angle from semg signals based on sca-lstm deep learning approach.
\newblock {\em Biomedical Signal Processing and Control}, 61:102024, 2020.

\bibitem{hajian2022deep}
Gelareh Hajian and Evelyn Morin.
\newblock Deep multi-scale fusion of convolutional neural networks for emg-based movement estimation.
\newblock {\em IEEE Transactions on Neural Systems and Rehabilitation Engineering}, 30:486--495, 2022.

\bibitem{sartori2016neural}
Massimo Sartori, David~G Llyod, and Dario Farina.
\newblock Neural data-driven musculoskeletal modeling for personalized neurorehabilitation technologies.
\newblock {\em IEEE transactions on biomedical engineering}, 63(5):879--893, 2016.

\bibitem{raissi2019physics}
Maziar Raissi, Paris Perdikaris, and George~E Karniadakis.
\newblock Physics-informed neural networks: A deep learning framework for solving forward and inverse problems involving nonlinear partial differential equations.
\newblock {\em Journal of Computational physics}, 378:686--707, 2019.

\bibitem{karniadakis2021physics}
George~Em Karniadakis, Ioannis~G Kevrekidis, Lu~Lu, Paris Perdikaris, Sifan Wang, and Liu Yang.
\newblock Physics-informed machine learning.
\newblock {\em Nature Reviews Physics}, 3(6):422--440, 2021.

\bibitem{zhang2022physics}
Jie Zhang, Yihui Zhao, Fergus Shone, Zhenhong Li, Alejandro~F Frangi, Sheng~Quan Xie, and Zhi-Qiang Zhang.
\newblock Physics-informed deep learning for musculoskeletal modeling: Predicting muscle forces and joint kinematics from surface emg.
\newblock {\em IEEE Transactions on Neural Systems and Rehabilitation Engineering}, 31:484--493, 2022.

\bibitem{kissas2020machine}
Georgios Kissas, Yibo Yang, Eileen Hwuang, Walter~R Witschey, John~A Detre, and Paris Perdikaris.
\newblock Machine learning in cardiovascular flows modeling: Predicting arterial blood pressure from non-invasive 4d flow mri data using physics-informed neural networks.
\newblock {\em Computer Methods in Applied Mechanics and Engineering}, 358:112623, 2020.

\bibitem{taneja2022feature}
Karan Taneja, Xiaolong He, QiZhi He, Xinlun Zhao, Yun-An Lin, Kenneth~J Loh, and Jiun-Shyan Chen.
\newblock A feature-encoded physics-informed parameter identification neural network for musculoskeletal systems.
\newblock {\em Journal of biomechanical engineering}, 144(12):121006, 2022.

\bibitem{london1981kinematics}
James~T London.
\newblock Kinematics of the elbow.
\newblock {\em JBJS}, 63(4):529--535, 1981.

\bibitem{holzbaur2005model}
Katherine~RS Holzbaur, Wendy~M Murray, and Scott~L Delp.
\newblock A model of the upper extremity for simulating musculoskeletal surgery and analyzing neuromuscular control.
\newblock {\em Annals of biomedical engineering}, 33:829--840, 2005.

\bibitem{herman2016physics}
Irving~P Herman.
\newblock {\em Physics of the human body}.
\newblock Springer, 2016.

\bibitem{hinrichs1985regression}
Richard~N Hinrichs.
\newblock Regression equations to predict segmental moments of inertia from anthropometric measurements: an extension of the data of chandler et al.(1975).
\newblock {\em Journal of Biomechanics}, 18(8):621--624, 1985.

\bibitem{konrad2005abc}
Peter Konrad.
\newblock The abc of emg.
\newblock {\em A practical introduction to kinesiological electromyography}, 1(2005):30--5, 2005.

\bibitem{sartori2012emg}
Massimo Sartori, Monica Reggiani, Dario Farina, and David~G Lloyd.
\newblock Emg-driven forward-dynamic estimation of muscle force and joint moment about multiple degrees of freedom in the human lower extremity.
\newblock {\em PloS one}, 7(12):e52618, 2012.

\bibitem{lloyd2003emg}
David~G Lloyd and Thor~F Besier.
\newblock An emg-driven musculoskeletal model to estimate muscle forces and knee joint moments in vivo.
\newblock {\em Journal of biomechanics}, 36(6):765--776, 2003.

\bibitem{durandau2019voluntary}
Guillaume Durandau, Dario Farina, Guillermo As{\'\i}n-Prieto, Iris Dimbwadyo-Terrer, Sergio Lerma-Lara, Jose~L Pons, Juan~C Moreno, and Massimo Sartori.
\newblock Voluntary control of wearable robotic exoskeletons by patients with paresis via neuromechanical modeling.
\newblock {\em Journal of neuroengineering and rehabilitation}, 16:1--18, 2019.

\end{thebibliography}

\end{document}